\documentclass[11pt,a4paper]{article}
\usepackage{epsfig}
\usepackage[T1]{fontenc}    
\usepackage{graphics}
\usepackage{graphicx}
\usepackage{pstricks,pst-coil,pst-fill,pst-plot}
\usepackage[fleqn]{amsmath}    
\usepackage{amssymb}    
\usepackage{amsfonts}   
\usepackage{verbatim}   
\usepackage{mathrsfs}   
\usepackage{dsfont}
\usepackage{euscript}
\usepackage{yfonts}
\usepackage{txfonts}
\usepackage{marvosym}
\usepackage{vmargin}        

\setmarginsrb{1.8cm}{2cm}{1.8cm}{2cm}{1cm}{1cm}{1cm}{1.6cm}
\makeatletter
\@addtoreset{equation}{section}
\makeatother

\let\tend=\rightarrow

\newtheorem{prop}{Proposition}[section]
\newtheorem{cor}{Corollary}[section]

\newtheorem{lemme}{Lemma}[section]

\def\Proof{\medskip\noindent {\it Proof --- \ }}

\let\qed=\cqfd

\newcommand\beq{\begin{equation}}
\newcommand\enq{\end{equation}}
\newcommand\bem{\begin{multline}}
\newcommand\enm{\end{multline}}

\def\beqa{\begin{eqnarray}}
\def\eeqa{\end{eqnarray}}
\def\ba{\begin{array}}
\def\ea{\end{array}}

\def\det{\operatorname{det}}

\newcommand{\f}[2]{{\ensuremath{%
    \mathchoice%
    {\dfrac{#1}{#2}}
    {\dfrac{#1}{#2}}
    {\frac{#1}{#2}}
    {\frac{#1}{#2}}
}}}
\newcommand{\tf}[2]{\ensuremath{#1/#2}}
\newcommand{\pa}[1]{\ensuremath{\left(#1\right)}}
\newcommand{\paa}[1]{\ensuremath{\left\{#1\right\}}}
\newcommand{\pac}[1]{\ensuremath{\left[#1\right]}}
\newcommand{\paf}[2]{\ensuremath{\left(\f{#1}{#2}\right)}}
\newcommand{\pab}[2]{\ensuremath{\pa{\ba{c} #1 \\ #2 \ea }}}
\newcommand{\pamb}[3]{\ensuremath{\pa{ #1 \left| \ba{c} #2 \\ #3 \ea \right. }}}

\def\a{\alpha}
\def\be{\beta}
\def\ga{\gamma}
\def\Ga{\Gamma}
\def\de{\delta}

\def\eps{\epsilon}

\def\la{\lambda}

\def\sg{\sigma}

\def\Sg{\Sigma}

\newcommand{\mc}[1]{\ensuremath{\mathcal{#1}}}
\newcommand{\mf}[1]{\ensuremath{\mathfrak{#1}}}

\newcommand{\ov}[1]{\ensuremath{\overline{#1}}}
\newcommand{\wt}[1]{\ensuremath{\widetilde{#1}}}

\newcommand{\Int}[2]{\ensuremath{\int\limits_{#1}^{#2}}}

\newcommand{\sul}[2]{\ensuremath{\sum\limits_{#1}^{#2}}}
\newcommand{\pl}[2]{\ensuremath{\prod\limits_{#1}^{#2}}}

\newcommand{\Dp}[1]{\ensuremath{\partial_{#1}}}

\newcommand{\ex}[1]{\ensuremath{\e{e}^{#1}}}

\newcommand{\ddet}[2]{\ensuremath{\det_{#1}\pac{#2}}}

\newcommand{\abs}[1]{\ensuremath{\left| #1 \right|}}

\newcommand{\moy}[1]{\ensuremath{\langle #1 \rangle}}

\newcommand{\dd}{\mathrm{d}}
\newcommand{\e}[1]{\ensuremath{\mathrm{#1}}}

\newcommand{\intff}[2]{\ensuremath{\left [ \, #1 \,; #2 \, \right ] }}

\begin{document}


\author{K.~K.~Kozlowski\footnote{ Laboratoire de Physique, UMR 5672 du CNRS,  ENS Lyon,  France,
 karol.kozlowski@ens-lyon.fr}}

\begin{flushright}
LPENSL-TH-03/09\\
\end{flushright}
\par \vskip .1in \noindent

\vspace{24pt}

\begin{center}
\begin{LARGE}
{\bf   Fine structure of the asymptotic expansion of cyclic integrals}
\end{LARGE}

\vspace{50pt}
\begin{large}
{\bf K.~K.~Kozlowski}\footnote[1]{ Laboratoire de Physique, ENS Lyon et CNRS,  France,
 karol.kozlowski@ens-lyon.fr}
\end{large}

\end{center}

\vspace{3cm}

\begin{abstract}
The asymptotic expansion of $n$-dimensional cyclic integrals was expressed as a series of functionals
acting on the symmetric function involved in the cyclic integral. In this article, we
give an explicit formula for the action of these functionals on a specific class of symmetric functions. These results are necessary for the computation of the
$\e{O}\pa{1}$ part in the long-distance asymptotic behavior of correlation functions in integrable models.
\end{abstract}

\newpage


\section{Introduction}

\hspace{3mm} There is a great interest in obtaining exact representations for the correlation functions of integrable models
in one-dimensional quantum mechanics\cite{JimMiwaMoriSatoSineKernelPVForBoseGaz,BoosJimboMiwaSmironovTakeyamaGrassMannStructureinXXZ,KMNTMasterEquation,KMTFormfactorsperiodicXXZ}. This is especially due to the recent development of experimental techniques allowing one to directly observe such models.
Among experimentally interesting correlation functions, one could mention the dynamical ones that are for instance measured through
neutron scattering on anti-ferromagnets. It was the determinant representation for the form factors of the spin-$\tf{1}{2}$ XXZ chain \cite{KMTFormfactorsperiodicXXZ} or the non-linear Schrodinger model \cite{KojimaKorepinSlavnovNLSEDeterminatFormFactorAndDualFieldTempeAndTime} that opened the possibility for an effective numerical study of these dynamical correlation functions in relatively long chains \cite{CauxHagemansMailletDynamicalCorrFunctXXZinFieldPlots} and diffuse lattice Bose gases \cite{CauxCalabreseSlavnovDynamicalCorrFunctBoseGasPlotsAndNumerics}
(this should be put in contrast with numerical diagonalization that is limited to twenty or so sites).
 One had to resort to numerics so as to obtain a satisfactory answer as, in general,
the exact representations for the correlation functions in integrable models are involved and often not computable in a closed form.
Even if simple closed expressions for the correlators do not exist, it is sometimes possible to built on these explicit answers so as to derive the asymptotic behavior of certain
correlation functions. This, in turn, allows one to test the conformal field theoretic predictions \cite{CardyConformalExponents} for the
power-law decay with the distance of two-point functions in gapless one-dimensional quantum Hamiltonians \cite{AffleckCFTStudyForGapless1DQuantHam,HaldaneCritExponentsAndSpectralProp}.
This last program has been carried recently by the author and his collaborators \cite{KozKitMailSlaTerXXZsgZsgZAsymptotics}. There, it was shown that it is possible to extract, starting from
first-principle based computations on the XXZ spin-$\tf{1}{2}$ chain, the long-distance $x$ power-law decay of the longitudinal spin-spin $\moy{\sg^z_1 \sg^z_{x}}$ correlation function. As expected, these asymptotics were found to match the conformal field theoretic predictions for this quantity.

We now briefly outline the method used for deriving these asymptotics. This will settle the context of the results presented in this paper. The method for extracting the asymptotic behavior of spin-spin correlation functions builds on the master equation \cite{KozKitMailSlaTer6VertexRMatrixMasterEquation,KMNTMasterEquation}. It is an $N$-fold contour integral representation for the generating function $\moy{\ex{\be \mc{Q}_x}}$ of two-point functions. Starting from this master equation, one should first reach the conformal regime of the model. The latter is supposed to manifest itself in the case of a gapless spectrum above the ground state. For integrable models, this corresponds to sending the size of the model to infinity and restricting oneself to a given range of the coupling constants. In particular, to reach this limit, one should send the number of integrals $N$ to infinity. However, the latter limit is impossible directly in the aforementioned integral representation.
To do so, one has first to expand the $N$-fold contour integral into a series, whose summand $a_n$ of order $n$ is a linear combination of $2n$-fold integrals. Each of these integrals depends on the large parameter $x$ and is expressed as products
of the so-called cyclic integrals:
\beq
\mc{I}_n\pac{\mc{F}_n}= \Int{ \Ga\pa{\intff{-q}{q}}  }{} \hspace{-3mm} \f{ \dd^n z  }{ \pa{2i\pi}^n  }   \Int{-q}{q}  \!  \f{ \dd^n \la}{ \pa{2i\pi}^n } \,
\mc{F}_n\pab{ \paa{\la} }{ \paa{z} }  \pl{k=1}{n}  \f{ \ex{ix \pac{ p\pa{z_k}-p\pa{\la_k}  }} }{ \pa{\la_k-z_k}\pa{\la_k-z_{k+1}}  } \; .
\label{definition integrale cyclique}
\enq
The functions $\mc{F}_n$ and $p$ are specific of the considered model.
The asymptotic behavior of \eqref{definition integrale cyclique} drives the one of $a_n$. It was shown in \cite{KozKitMailSlaTerRHPapproachtoSuperSineKernel}
that the asymptotic expansion (AE) of cyclic integrals can be deduced from the asymptotic solution of a Riemann--Hilbert problem related to the generalized sine kernel. This AE is written as a series of functionals $I^{\pa{N,m}}_n$ acting on the function $\mc{F}_n$.
In order to deduce the AE of
$\moy{\ex{\be \mc{Q}_x}}$, one should replace every cyclic integral in $a_n$ by its AE and then re-order an re-sum the resulting terms. Due to the
product structure of $a_n$ such a procedure involves summing-up series of products of functionals $I^{\pa{N,m}}_n$.
In the intermediate steps of this summation procedure, it happens that the $x$-independent function one started with, acquires an $x$-dependence.
It thus follows that one has to know how to evaluate the action of the functionals $I_n^{\pa{N,m}}$ on certain classes of $x$-dependent functions.
This last topic constitutes the subject of this article.
It seems rather difficult to give an explicit formula for  $I_n^{\pa{N,m}} \pac{\mc{F}_n}$ when $\mc{F}_n$ is a generic symmetric function.
Indeed, the functionals $I_n^{\pa{N,m}}$ have been built  rather implicitly through a recursive procedure.
However, we found that it is possible to write a simple explicit formula for the leading order in $x$ part of the action of $I_n^{\pa{N,m}}$
when focusing on a specific class of $x$-dependent functions.
Such a result should not be understood as some part of the asymptotics of $2n$-fold cyclic integrals versus some $x$-dependent functions but rather as the action of given functionals (defined as the ones appearing in the AE of cyclic integrals) on a particular kind of functions.
In this paper, we only discuss the leading order in $x$ part of the action although it is quite clear how to apply the method we present so as to obtain further, sub-leading, corrections.
Of course, the size and complexity of the formulae will grow very quickly with the precision of the corrections

The article is organized as follows. In section 2 we remind some properties of the AE of the generalized sine kernel which is a building block
of cyclic integrals \eqref{definition integrale cyclique}. We prove some results on the structure of these asymptotics. In section 3, we recall the link between the generalized sine kernel and cyclic integrals. We also remind the overall properties of their AE and derive some sum-rules.
Finally, in section 4, we apply these sum rules so as to determine explicitly the leading $\e{O}\pa{1}$ part of the action of these functionals
on a certain class of $x$-dependent functions.


\section{The generalized sine kernel}

The generalized sine kernel is an integral operator $I+V$ on $L^2\pa{ \intff{-q}{q} }$. Its kernel $V$ reads
\beq
V\pa{\la,\mu}= \ga \sqrt{F\pa{\la}F\pa{\mu}} \f{ \tf{e\pa{\la}}{e\pa{\mu}}- \tf{e\pa{\mu}}{e\pa{\la}} }{ 2i \pi \pa{\la-\mu} }
 \;\; , \quad e\pa{\la}= \ex{i \tf{x p\pa{\la}}{2} + \tf{g\pa{\la}}{2} } \; .
\label{definition of the GSK}
\enq
The asymptotic behavior of the Fredholm determinant of this kernel was derived in \cite{KozKitMailSlaTerRHPapproachtoSuperSineKernel} under
the assumption of   $p,F, g$  holomorphic in a neighborhood of $\intff{-q}{q}$, $p$ injective on this neighborhood and $\abs{\ga}$ small enough.
We remind that the AE of its logarithm has the structure
\beq
\ln\ddet{}{ I+V } = \sul{N=-1}{+\infty} \f{1}{x^N} \mc{R}^{\pa{N,0}}\pac{p,g, \nu } +
\sul{m \in \mathbb{Z}^*}{}\sul{N \geq 2\abs{m} }{} \f{1}{x^N} \mc{R}^{\pa{N, m}} \pac{p,g, \nu }  \ex{im x\pa{p_+-p_-}} x^{m\pa{\nu_+ + \nu_-}} \; .
\label{Asymptotic expansion log Fredholm}
\enq
Here and in the following, we use the notation $H_{\pm}=H\pa{\pm q}$ for the boundary values of any function $H$.
The function $\nu$ is related to $F$ by
\beq
\nu\pa{\la}=\f{i}{2\pi} \ln \pa{1+\ga F\pa{\la}} \; .
\label{definition fonction nu}
\enq
and  $\mc{R}^{ \pa{N,m} }\pac{p,g, \nu }$, $m\in \mathbb{Z}$, are functional in $p, g$ and $\nu$.
They are also  polynomials in $\ln x$ of degree $N$ (we decided not to insist explicitly on this dependence in order to keep the notation light). $\mc{R}^{\pa{N}}$ is expressible in terms of integrals over $\intff{-q}{q}$
involving $\nu$ and products of functions in $p_{\pm}, g_{\pm}, \nu_{\pm}$ or of their higher order derivatives.

Actually, it is easy to see from the details of the recursive construction of $\mc{R}^{\pa{N,m}}$ given in \cite{KozKitMailSlaTerRHPapproachtoSuperSineKernel} that these functionals only produce algebraic expressions in $p$
or its derivatives. Moreover, only $p'_{\pm}$ or $p_+-p_-$ appear in the denominator of such expressions\footnote{ Another argument
to convince oneself of this fact is that, otherwise, the AE would be ill-defined on the functions fulfilling the assumptions of \cite{KozKitMailSlaTerRHPapproachtoSuperSineKernel}.
Indeed, the AE is obtained to any order in $x$ only by assuming holomorphy of the involved functions, smallness of $\abs{\ga}$ and
injectivity of $p$. These conditions can only guarantee that $p_+-p_-$ and $p'_{\pm}$ do not vanish.}.

Note that one can represent the operator $I+V$ in several ways. Namely, the kernels  $V_{p,F,g}$ and $V_{p-i\tf{g}{x},F,0}$
parameterize the same generalized sine kernel\footnote{Here we have explicitly stressed out the dependence of $V$ on the functions $p,F,g$}. However, these two different parameterizations lead to, a priori distinct, asymptotic series\footnote{We stress that one can also apply the results of \cite{KozKitMailSlaTerRHPapproachtoSuperSineKernel}
in the case of an $x$-dependent $p$ function, as long as $p=p_0+\e{o}\pa{1}$ and
$p_0$ is holomorphic in a neighborhood of $\intff{-q}{q}$ and injective on this neighborhood. This stems from the fact that when $x\tend +\infty$, only $p_0$ plays a role in the details of the analysis carried in \cite{KozKitMailSlaTerRHPapproachtoSuperSineKernel}.} \eqref{Asymptotic expansion log Fredholm}. In virtue of the uniqueness of an asymptotic expansions, these
have to coincide
\bem
\sul{N=-1}{+\infty} \f{1}{x^N} \mc{R}^{\pa{N,0}}\pac{p,g, \nu } +
\sul{m \in \mathbb{Z}^*}{}\sul{N \geq 2\abs{m} }{} \f{1}{x^N} \mc{R}^{\pa{N, m}} \pac{p,g, \nu }  \ex{im x\pa{p_+-p_-}}
x^{m\pa{\nu_+ + \nu_-}} \;  \sim \\
\; \sul{N=-1}{+\infty} \f{1}{x^N} \mc{R}^{\pa{N,0}}\pac{p-i\tf{g}{x},0, \nu } +
\sul{m \in \mathbb{Z}^*}{}\sul{N \geq 2\abs{m} }{} \f{1}{x^N} \mc{R}^{\pa{N, m}} \pac{p-i\tf{g}{x},0, \nu }  \ex{imx \pa{p_+-p_-}}
\ex{m\pa{g_+-g_-}} x^{m\pa{\nu_+ + \nu_-}} \; .
\label{Egalite DA}
\end{multline}
Due to the aforementioned properties of the functionals $\mc{R}^{\pa{N,m}}$, the dependence in the RHS of \eqref{Egalite DA} is algebraic in $p-i\tf{g}{x}$ and the only terms appear in the denominator have a well defined Taylor series ($ie$ one that only produces corrections in respect to the $x \tend + \infty$ limit). Hence, $\mc{R}^{\pa{N}} \pac{p-i\tf{g}{x},0, \nu }$ can be
expanded in a Taylor series of inverse powers of $x$.
Such an expansion yields
$\mc{R}^{\pa{N}}\pac{p,0, \nu }$ and terms that involve $g$ but are subdominant with respect to it.
In other words, the structure in $g$ of the  RHS of \eqref{Egalite DA}  follows
by reordering and gathering the powers in $x$ issued from the various Taylor series expansions in the LHS of \eqref{Egalite DA}. The mechanism of this effect can be seen explicitly on the leading asymptotics of
$\ln \ddet{}{I+V_{p,g,F}}$:
\beq
x \mc{R}^{\pa{-1,0}} \pac{p,0, \nu } = x \Int{-q}{q} p'\pa{\la} \nu\pa{\la} \dd \la \; .
\enq
The substitution $p \mapsto p-i\tf{g}{x}$ produces
\beq
x \Int{-q}{q} p'\pa{\la} \nu\pa{\la} \dd \la -i \Int{-q}{q} g'\pa{\la} \nu\pa{\la} \dd \la \; .
\label{Exemple apparition g terme lineaire}
\enq
This additional $g$-dependent term is precisely the one appearing in the $\e{O}\pa{1}$ asymptotics of $\ln\ddet{}{I+V_{p,F,g}}$, \textit{cf} \cite{KozKitMailSlaTerRHPapproachtoSuperSineKernel}.

\subsection{Some results on the structure of the asymptotic series}
One can build on the two  ways of presenting the AE so as to
deduce some properties on the structure of the coefficients involved in the AE \eqref{Asymptotic expansion log Fredholm}. Such results can hardly
by inferred from a direct computation based on the techniques presented in \cite{KozKitMailSlaTerRHPapproachtoSuperSineKernel}.
We prove the results that are important for the purpose of the further analysis in the two lemmas below.
\begin{lemme}
\label{lemme terme non-oscillant en g}
The $\tf{1}{x^N}$ order in the AE of $\ln\ddet{}{I+V}$ contains the term
\beq
\sul{\sg=\pm}{}  \f{1}{N} \paf{ i g'_{\sg} }{ p'_{\sg} }^N  \nu^2_{\sg}  \; .
\label{expression term non-oscillant en g a la N}
\enq
This is the highest possible power of $g$ that can appear in the non-oscillating part of the $x^{-N}$ order in the AE of $\ln\ddet{}{I+V}$ \eqref{Asymptotic expansion log Fredholm}.
\end{lemme}

\Proof
As it was discussed previously, it is possible to deduce the whole $g$-dependence of $\mc{R}^{\pa{N,0}}\pac{p,g,\nu}$
by expanding $\mc{R}^{\pa{0,0}}\pac{p-i\tf{g}{x}, 0 ,\nu}, \dots, x^{1-N}\mc{R}^{\pa{N-1,0}}\pac{p-i\tf{g}{x}, 0 ,\nu}$
in to Taylor series of inverse powers of $x$. Note that the $g$-dependent part of $x^{-k}\mc{R}^{\pa{k,0}}\pac{p-i\tf{g}{x}, 0 ,\nu}$, $k \geq N$
only produces subdominant contribution with respect to $x^{-N}\mc{R}^{\pa{N}}$.
Moreover, we did not include $\mc{R}^{ \pa{-1,0}}$ in the list, as we have already seen \eqref{Exemple apparition g terme lineaire} that it only generates the $g$ dependent part of $\mc{R}^{\pa{0}}\pac{p,g,\nu}$.

The Taylor expansion into inverse powers of $g$ of the functional $R^{\pa{\ell,0}}\pac{p-i\tf{g}{x},0,\nu}$ is of the type
\beq
R^{\pa{\ell,0}}\pac{p,0,\nu} + \sul{n \geq 1}{} x^{-n}F_n^{\pa{\ell} }\pac{p,g,\nu} \; .
\enq
The functional $F^{\pa{\ell}}_n\pac{p,g,\nu}$ is homogeneous
in $g$ of order $n$, $\textit{ie}$ $F^{\pa{p}}_n\pac{p,t g,\nu}= t^nF^{\pa{p}}_n\pac{p,g,\nu}$. In other
words it only contains powers of $g$ whose total degree in $g$ is $n$, for instance terms of the type
\beq
\pa{g_+}^{n_+^{\pa{0}}}\dots\pa{ g^{\pa{k}}_+ }^{n_+^{\pa{k}}}\pa{g_-}^{n_-^{\pa{0}}}\dots\pa{ g^{\pa{k}}_- }^{n_-^{\pa{k}}} \; ,
 \qquad \e{where} \quad  \sul{p=1}{k} n_+^{\pa{p}}+n_-^{\pa{p}}=n.
\enq
 Hence, the highest possible degree in $g$ of
$\mc{R}^{\pa{N,0}}\pac{p,g,\nu}$ can only be issued from the $N^{th}$ term of the Taylor series in $x^{-1}$
of $\mc{R}^{\pa{0}}\pac{p-i\tf{g}{x},0,\nu}$. As $- \Sg_{\sg=\pm} \nu_{\sg}^2 \ln p'_{\sg}$ represents
the $p$ dependent part of $\mc{R}^{\pa{0}}\pac{p,0,\nu}$, the substitution $p \mapsto p-i\tf{g}{x}$ generates the series
of inverse powers of $x$:
\beq
-\sul{\sg=\pm}{} \nu^2_{\sg} \ln p_{\sg}' +  \sul{\sg=\pm}{} \nu^2_{\sg} \sul{\ell \geq 1}{}  \f{1}{\ell \, x^{\ell}}\paf{i \, g'_{\sg}}{p'_{\sg}}^{\ell} \; .
\enq
Thus, the highest possible degree in $g$ contained in $\mc{R}^{\pa{N,0}}\pac{p,g,\nu}$ is $N$ and
the corresponding contribution is
$ N^{-1}\cdot  \pac{ \pa{ \tf{ i g'_{+}}{p'_{+}} }^{N} +  \pa{ \tf{i g'_{-}}{p'_{-}} }^{N}} $.  \qed

One can also prove an analogous lemma in respect to the oscillating part.
\begin{lemme}
\label{lemme term oscillant en g a la n}
The $\tf{1}{x^{N+2}}$ part of the oscillating corrections with period $\ex{ i \sg x \pa{p_+-p_-}}$, $\sg=\pm$, contains the series
\beq
\ex{\sg \pa{g_+-g_-}} \sul{k=0}{n} \paf{i g_+' }{p'_+}^k \paf{i g_-' }{p'_-}^{n-k}
 \wt{O}_{n,\, k}\pac{ \sg \, \nu}  \; .
\label{expression term oscillant ordre n}
\enq
We have introduced the functional
\beqa
\wt{O}_{n,\, k}\pac{\nu} &=& \f{  \pac{2q x}^{2\pa{\nu_+ + \nu_-}}  }{ \pa{2q}^2 }
{p'_+}^{2\nu_+-1} {p'_-}^{2 \nu_- -1 }
\f{ \Ga\pa{1-\nu_-} \Ga\pa{1-\nu_+} \kappa^2_{-}}{ \Ga\pa{\nu_-} \Ga\pa{\nu_+}  \kappa_+^2 }
\f{ \pa{2 \nu_+-1}_k \pa{2\nu_{-}-1}_{n-k} }{ k! \pa{n-k}!} \; \; ,  \\
 \e{with} \; \; \ln\kappa\pa{\la} & =&  \Int{-q}{q} \f{\nu\pa{\la}-\nu\pa{\mu}}{\la-\mu} \dd \mu \;.
\eeqa
\end{lemme}

\Proof
The first oscillating term appearing in the asymptotic expansion of $\ln \ddet{}{I+V} _{\mid_{g=0}}$
reads \cite{KozKitMailSlaTerRHPapproachtoSuperSineKernel}:
\beq
\f{1}{x^2} \sul{\sg=\pm}{} \wt{O}_{0,0}\pac{\sg \nu} \ex{ix \sg \pa{ p_+ -p_-}}  \;
\quad \e{so}\, \e{that} \quad R^{\pa{2,\sg}}\pac{p,g,\nu}= x^{-2 \sg \pa{\nu_+ + \nu_-} } \wt{O}_{0,0}\pac{\sg \nu}
\enq
Just as for the non-oscillating corrections discussed above, it is the only one responsible for producing the highest degree in $g$
of the oscillating corrections subordinate to the frequency $\ex{ix \sg \pa{p_+-p_-}}$, $\sg=\pm$. To deduce the structure of these terms, we set
$p \mapsto p-i\tf{g}{x}$ in the above functional and perform a Taylor series expansion. For $x$ large enough:
\beq
\f{1}{\pa{1-\tf{\eps_1}{x}}^{\a}\pa{1-\tf{\eps_2}{x}}^{\be} }= \sul{n \geq 0}{} \f{1}{x^{n} } \sul{k=0}{n} \eps_2^{n-k} \eps_1^k
\f{ \pa{\a}_k \pa{\be}_{n-k} }{k! \pa{n-k}! } \; , \qquad  \e{where} \quad   \pa{\a}_k = \f{ \Ga\pa{\a+k} }{\Ga\pa{\a }  } \; .
\enq
We get that the highest degree in $g$ appearing  in the $x^{-2-N}$ part of the oscillating $\ex{i\sg\pa{p_+-p_-}}$ corrections
is given by
\beq
\sul{ \sg = \pm }{}\sul{k=0}{N}  \wt{O}_{N,k}\pac{ \sg \nu } \ex{\sg \pa{g_+-g_-}}  \paf{i g_+' }{p'_+}^k \paf{i g_-' }{p'_-}^{N-k} \;.
\enq
\qed

\vspace{5mm}
It is clear that one can look at lower order terms like $\mc{R}^{ \pa{1,0}} \pac{p,0,\nu}$ so as to obtain
the coefficient in front of the terms of total degree $N-1$ in $g$ at any order in $x^{-N}$.
In the next sections, we will actually need the highest degree in $g$ dependence in the AE of
\beq
J_n\pac{p,g,F} = \f{ \pa{-1}^{n-1} }{ \pa{n-1}! } \Dp{\ga}^n \ln \ddet{}{I+V} _{\mid_{\ga=0}} \; .
\enq
The latter is easily deduced from the previous lemmas due to the fact that the AE of $\ln \ddet{}{I+V}$
in uniform to any finite order derivative in $\ga$. We gather these results in the

\begin{cor}
The highest degree of $g$ in the non-oscillating $x^{-N}$ order of the
asymptotic expansion of $J_n\pac{p,g,F}$ reads
\beq
\f{\pa{-1}^{n-1}}{\pa{n-1}!} \sul{\sg=\pm}{}  \f{1}{N} \paf{ i g'_{\sg} }{ p'_{\sg} }^N  \pa{F_{\sg} }^n {  \Dp{\ga}^n \pa{\nu^2_{0}}  }_{\mid_{\ga=0} }
\quad \e{where}  \quad \nu_0= \f{i}{2\pi} \ln \pa{1+\ga} \; .
\enq
Similarly, the highest degree in $g$ present in the $x^{-N-2}$ oscillating part at frequency $\ex{ix \sg \pa{p_+-p_-} }$, $\sg=\pm$,
in the AE of $J_n\pac{p,g,F}$ is given by the series
\beq
 \f{ \pa{-1}^{n-1} }{ \pa{n-1}!} \ex{\sg \pa{g_+-g_-}} \sul{k=0}{N} \paf{i g_+' }{p'_+}^k \paf{i g_-' }{p'_-}^{N-k}
\Dp{\ga}^n \wt{O}_{N,\, k}\pac{\sg \nu}_{\mid_{\ga=0} }  \; .
\enq
\end{cor}

\section{Cyclic multiple integrals}
One of the main results of \cite{KozKitMailSlaTerRHPapproachtoSuperSineKernel}, was the proof of the asymptotic expansion for
cyclic integrals:
\beq
\mc{I}_n\pac{\mc{F}_n}= \Int{ \Ga\pa{\intff{-q}{q}}  }{} \hspace{-3mm} \f{ \dd^n z  }{ \pa{2i\pi}^n  }   \Int{-q}{q}  \!  \f{ \dd^n \la}{ \pa{2i\pi}^n } \,
\mc{F}_n\pab{ \paa{\la} }{ \paa{z} }  \pl{k=1}{n}  \f{ \ex{ix \pac{ p\pa{z_k}-p\pa{\la_k}  }} }{ \pa{\la_k-z_k}\pa{\la_k-z_{k+1}}  } \; .
\enq
%
%
There, $\mc{F}_n$ is a holomorphic function of the $n$ variables $\paa{z}$ and the $n$ variables $\paa{\la}$ that is symmetric separately in respect to variables belonging to each of these two sets. $\Ga\pa{\intff{-q}{q}}$ is a sufficiently small loop of index $1$ around $\intff{-q}{q}$.
Lastly, we agree upon $z_{n+1}\equiv z_1$. It was shown in \cite{KozKitMailSlaTerRHPapproachtoSuperSineKernel}, that there exists functionals
$I_n^{\pa{N,m}}$ such that the asymptotic expansion of $\mc{I}_n\pac{\mc{F}_n}$ reads
\beq
\mc{I}_n\pac{ \mc{F}_n } =    \sul{ m\in \mathbb{Z}   }{} \sul{N \geq 2\abs{m} -\de_{m,0}}{}
                            \ex{i m x \pa{p_+-p_-}}  \f{1}{x^N}I_n^{ \pa{N,m} }\pac{\mc{F}_n} \; .
\label{Asymptotic expansion cyclic integral}
\enq
This equality is to be understood in the sense of an asymptotic expansion. In particular, it excludes any $\e{O}\pa{x^{-\infty}}$ terms.
For the purpose of our study, we now recall a couple of general properties of the functionals appearing in \eqref{Asymptotic expansion cyclic integral}. First, all the oscillating terms $\ex{imx \pa{p_+-p_-}}$ are factored out from $I_{n}^{\pa{N,m}}$ and these functionals
are polynomial in $\ln x$ of at most degree $n+N$. Their action involves the finite-difference operator
\beq
\eth^{\pa{t}}_{z_{\ell}}\pa{\mu} \cdot \mc{G}_n\pac{ \paa{ \paa{z_{i}}^{\bar{p}_i } }_{i=1,\ldots,r} }
=
\mc{G}_n\pac{ \paa{ \paa{z_{i}}^{\bar{p}_i } }_{i=1,\ldots,r} }
-
\mc{G}_n\pac{ \paa{ \paa{z_{k}}^{\bar{p}_k } }_{k\not= \ell} , \paa{\mu}^{t}, \paa{z_{\ell}}^{\bar{p}_\ell-m}} \, .
\label{definition operateur de difference finie}
\enq
The function $\mc{G}_n$ appearing in \eqref{definition operateur de difference finie} is assumed symmetric in its variables.
Moreover the notation, $\paa{z_k}^{\ov{p}_k}$ means that the variable $z_k$ is repeated $\ov{p}_k$ times
\beq
\paa{z_k}^{\ov{p}_k}=( \underbrace{z_1, \dots, z_1}_{\ov{p}_1 \, \e{times}}, \dots,  \underbrace{z_r,\dots z_r}_{ \ov{p}_r }   ) \; .
\enq
Hence, $\eth^{\pa{t}}_{z_{\ell}}\pa{\mu}$ acts by replacing $t\leq \ov{p}_{\ell}$ variables $z_{\ell}$ by $\mu$ and then performing the difference
with the function one started with. The action of $I_n^{\pa{N,m}}$ takes the form
\beq
I_n^{ \pa{N,m} }\pac{\mc{F}_n}=   \mf{S}^{\pa{m}}_N \,  * \!\! \sul{ \Sg k_{\ell} \leq  N-2\abs{m} }{}  G^{ \mf{S}^{\pa{m}}_N  }_{\paa{k_{\ell}}}
 \cdot \paa{  \Dp{z_1}^{k_1} \dots \Dp{z_r}^{k_r}  \pl{\ell=1}{r} \pl{t=1}{n} \pl{j=1}{p_{\ell\, , t}}
\paa{   \Int{-q}{q}  \f{\dd \mu_{\ell, \, t, \, j} }{z_{\ell}-  \mu_{\ell, \, t, \, j} } \eth^{\pa{t}}_{z_{\ell}}\pa{\mu_{\ell, \, t, \, j}} }
\cdot \mc{F}_n \pab{ \paa{z_{\ell}}^{\ov{p}_{\ell}} }{ \paa{z_{\ell}}^{\ov{p}_{\ell} + \eps_{\ell}}  }
  }_{z_{\ell}=\sg_{\ell}q}   \hspace{-3mm} .
\label{definition Functional ImNm}
\enq
Here $\mf{S}^{\pa{m}}_N * $ is a shorthand notation for the multiple summation symbol
\beq
\mf{S}^{\pa{m}}_N  *= \sul{r=1+\abs{m}}{n} \sul{\sg_{\ell}=\pm }{} \sul{ \substack{\eps_{\ell}=\pm 1, 0 \\ \Sg \eps_{\ell}=0}  }{}
\sul{ \substack{ \Sg \ov{p}_{\ell} = n  \\ p_{\ell,\, 0} \geq 0 }}{}
\label{definition somme Sigma}
\enq
over the points $\sg_{\ell}q=$, $\sg_{\ell}=\pm$, where the variables $z_{\ell}$
are evaluated, the
integers $\eps_{\ell}=\pm1 ,\,  0$, $\ell=1,\dots r$ parameterizing the sets of unequal variables and integers $p_{\ell,t}$ labeling the number of finite difference operators involved. These integers are subject to the  constraints $p_{\ell,t}\geq 0$, $p_{\ell,0} \geq 1$ and $\ov{p}_{\ell}=\Sg_{t=1}^{n} t p_{\ell,t} \; $, $\; \Sg_{\ell=1}^{r} \eps_{\ell}=0$ and $\Sg_{\ell=1}^{r} \eps_{\ell} p_{\sg_{\ell}}=m \pa{p_+-p_-}$.
The coefficients $G^{\mf{S}^{\pa{m}}_N}_{\paa{k_{\ell}}}$ depend on all of these additional summation indices and are polynomials in $\ln x$
of degree $n + N$. We chose to keep this dependence
implicit, so as to have as light notations as possible.

$I_n^{ \pa{N;m}}$ is thus written as a sum of terms. Each of these terms corresponds to several operation performed on $\mc{F}_n$. The set the initial $\la$-variables
of $\mc{F}_n$ is set to $\paa{z_{\ell}}^{\ov{p}_{\ell}}$ and the $z$-variables to $\paa{z_{\ell}}^{\ov{p}_{\ell}+\eps_{\ell}}$.
It thus follows that most of the variables in the upper and lower   hypergeometric type notation  for $\mc{F}_n$ are set equal. The variables $\paa{z_{\ell}}$ are acted upon by the finite-difference operators, the result of the action being integrated versus a $z_{\ell}$ dependent weight along $\intff{-q}{q}$.
After the action of the finite-difference operators, one acts with the derivative operators
$\Dp{z_1}^{k_1}\dots \Dp{z_r}^{k_r}$ with $0\leq \Sg_{s=1}^{r} k_s \leq N-2\abs{m}$ and sets $z_{\ell}=\sg_{\ell} q$, $\sg_{\ell}=\pm $.
 One can be more specific about the structure of the part
of $I_n^{\pa{N,m}}$ containing only the highest order $N-2\abs{m}$ of derivatives. There, the variables $z_1,\dots, z_r$
are set equal to $\sg_s q=z_s$, where the sequence $\sg_s \in \paa{\pm 1}$ and has precisely $\abs{m}$ jumps.
For instance, when $m=0$, only two sequences $\sg_{\ell}$ are possible, $\sg_s=\sg \, \forall s$, $\sg=\pm$.

There exist a relation between the Fredholm determinant $\ln \ddet{}{ I+V}$  and  cyclic integrals
involving a special class of of symmetric functions in $n$ variables $\paa{\la}$ and $n$ variables $\paa{z}$
that we call pure product functions:
\beq
\mc{F}^{\pa{\varphi,g}}_n \pab{ \paa{\la} }{ \paa{z} } = \pl{k=1}{n} \varphi\pa{\la_k} \ex{g\pa{z_k}} \; .
\enq
Indeed, on the one hand by looking at the $n^{\e{th}}$ term in the series in $\ga$ of $\ln \ddet{}{I+V}$ and, on the other hand, by computing the residues at $z_k=\la_k$ or $z_k=\la_{k+1}$
in \eqref{definition integrale cyclique}, one shows that
\beq
J_n\pac{p,g, \varphi \ex{g}} = \mc{I}_n\pac{ \mc{F}_n^{\pa{\varphi,g}} } \; .
\label{Link Cyclic Integral And Jn}
\enq
This observation was used in \cite{KozKitMailSlaTerRHPapproachtoSuperSineKernel} to build on the AE of $\ln\ddet{}{I+V}$
so as to deduce the one of general cyclic integrals \eqref{definition integrale cyclique}. Due to \eqref{Asymptotic expansion log Fredholm} and \eqref{Link Cyclic Integral And Jn} the $g$-dependent part of the AE of $\mc{I}_n\pac{ \mc{F}_n^{\pa{\varphi,g}} }$
can be deduced from the one of $\mc{R}^{\pa{N,m}}\pa{x}\pac{p,g,\nu^{\pa{\varphi\ex{g}}} }$, where $\nu^{\pa{\varphi\ex{g}}}$ is obtained from $\nu$
by setting $F=\varphi \ex{g}$ in \eqref{definition fonction nu}. Yet, this $g$-dependence has also to be reproduced if one chooses to
act with the functionals $I_n^{\pa{N,m}}$ \eqref{Asymptotic expansion cyclic integral} on $\mc{F}_n^{\pa{\varphi,g}} $ instead. This fact allows us to derive certain sum-rules for part of the terms appearing in \eqref{definition Functional ImNm}. Such identities will be used in the next section so as to obtain the leading $\e{O}\pa{1}$ order in $x$ of the action of $x^{-N} I_{n}^{\pa{N,m}}$ on $x$-dependent functions.

\subsection{Non-oscillating sum-rules}
We identify the part of $I_n^{\pa{N,0}}\pac{ \mc{F}_n^{\pa{\varphi,g}} }$ producing the highest order in $g$ of $J_n\pac{p,g, \varphi \ex{g}}$.
Due to the pure-product structure of $\mc{F}_n^{\pa{\varphi,g}} $ we have
\beq
\mc{F}^{\pa{\varphi,g}}_n\pab{ \paa{z_{\ell}}^{\ov{p}_{\ell}} }{ \paa{z_{\ell}}^{ \ov{p}_{\ell} + \eps_{\ell} } }  =
\pl{s=1}{r}F\pa{z_s}^{\ov{p}_s} \ex{\eps_s g\pa{z_s}} \; \; , \quad F=\varphi \ex{g} \; .
\enq
This function admits a simple representation for the action of
$\eth^{\pa{t}}_{z}\pa{\mu_{\ell, \, t, \, j}} $:
\bem
\pl{\ell=1}{r} \pl{t=1}{n} \pl{j=1}{p_{\ell\, , t}}
\paa{   \Int{-q}{q}  \f{\dd \mu_{\ell, \, t, \, j} }{z_{\ell}-  \mu_{\ell, \, t, \, j} } \eth^{\pa{t}}_{z_{\ell}}\pa{\mu_{\ell, \, t, \, j}} }
\cdot \pl{s=1}{r} \paa{  F\pa{z_s}^{\ov{p}_s} \ex{\eps_s g\pa{z_s}}  } = \\
 \pl{s=1}{r} \paa{   F\pa{z_s}^{p_{s,0}} \ex{\eps_s g\pa{z_s}}
\pl{t=1}{n}\pac{   \Int{-q}{q}  \f{ \pac{F\pa{z_s}}^{t}-\pac{F\pa{\mu}}^{t} }{z_{s}-  \mu } \dd \mu  }^{p_{s \, t}}  }\; .
\end{multline}
Thence,
\beq
I_n^{ \pa{N,0} }\pac{\mc{F}_n}=  \mf{S}^{\pa{0}}_N \,  * \!\! \sul{ \Sg k_{\ell} \leq N }{}
G^{ \mf{S}^{\pa{0}}_N }_{ \paa{k_{\ell}} }  \paa{
\Dp{z_1}^{k_1} \dots \Dp{z_r}^{k_r}
\pl{s=1}{r} \paa{F\pa{z_s}^{p_{s,0}} \ex{\eps_s g\pa{z_s}} }
\pl{t=1}{n}\paa{   \Int{-q}{q}  \f{ \pac{F\pa{z_s}}^{t}-\pac{F\pa{\mu}}^{t} }{z_{s}-  \mu } \dd \mu  }^{p_{s \, t}}
  }
\enq
The maximal degree in $g$ can only stem from the part of the sum
involving the highest possible number of derivatives. Indeed, it is obtained by
hitting with all of the derivatives $\Dp{z_1}^{k_1}\dots \Dp{z_r}^{k_r}$ on the exponent $\prod_{k=1}^{r} \ex{\eps_k g\pa{z_k}}$.
Any other type of action produces a lower degree in $g$. More explicitly, for an holomorphic function $H$
\beq
\Dp{z_1}^{k_1} \dots \Dp{z_r}^{k_r}  \paa{  \pl{s=1}{r} \ex{\eps_s g\pa{z_s}}  \cdot H\pa{ \paa{z}}}  =
 \pl{s=1}{r}  \pa{\eps_s g'\pa{z_s}}^{k_s} \ex{\eps_s g\pa{z_s}}  \cdot H\pa{ \paa{z}} + \e{LD}\pa{g} \; \; .
\label{action derivée multiple fonction exposant}
\enq
$\e{LD}\pa{g}$ stands for the terms that are of a lower degree in $g$. Once the derivatives have been acted with, one should set $z_1=\dots=z_r=\sg q$ as there are no jumps in the sequence $\sg_{\ell}$ when $m=0$ and $\Sg k_{\ell}=N$. In particular, this allows us
to simplify the products of the exponents
\beq
\pl{s=1}{r} \ex{ \eps_s g\pa{\sg q}} = \exp\paa{g\pa{ \sg q} \sul{s=1}{r} \eps_s  }=1 \quad \e{due}\,\e{to} \, \e{the} \, \e{constraint} \, \quad
\sul{s=1}{r} \eps_s=0 \; \; .
\enq

By lemma \ref{lemme terme non-oscillant en g}, the highest possible degree in $g$ appearing in the non-oscillating $x^{-N}$
order of the AE of $\ln\ddet{}{ I+V}$ is given by \eqref{expression term non-oscillant en g a la N}. Thus we obtain the sum-rule:
\bem
I_n^{ \pa{N,0} }\pac{\mc{F}^{\pa{\varphi,g}}_n}=
\mf{S}^{\pa{0}}_N *
\sul{ \Sg k_{\ell} = N }{}
G^{ \mf{S}^{\pa{0}}_N }_{  \paa{k_{\ell}} }  \pl{s=1}{r}  \paa{
 \pac{ \eps_s g'\pa{z_s}}^{k_s}  \pac{F\pa{z_s}  }^{p_{s,0}}
\pl{t=1}{n}\paa{   \Int{-q}{q}  \f{ \pac{ F\pa{z_s} }^t -\pac{ F\pa{\mu} }^t  }{z_{s}-  \mu } \dd \mu  }^{p_{s \, t}}
  } \\
=
\sul{\sg={\pm}}{ } \f{1}{N} \paf{i g'_{\sg} }{ p'_{\sg } } ^{N} \paa{ \f{\pa{-1}^{n-1}}{\pa{n-1}!} \Dp{\ga}^n \pa{\nu_0}_{\mid_{\ga=0}}  }
F^n_{\sg }  \;\; .
\label{sum-rule non-oscillating}
\end{multline}

\subsection{Oscillating sum-rules}
One can repeat word for word the above analysis in the case of the action of $I_{n}^{\pa{N,m}}$, $m=\pm $, on pure product functions.
One gets that
\bem
\mf{S}^{\pa{m}}_N  \, * \!\! \sul{ \Sg k_{\ell} = N-2 }{}
G^{ \mf{S}^{\pa{m}}_N }_{\paa{k_{\ell}}}   \pl{s=1}{r} \paa{
 \pac{\eps_s g'\pa{z_s}}^{k_s} \pac{F\pa{z_s} }^{p_{s,0}}
\pl{t=1}{n}\paa{   \Int{-q}{q}  \f{   \pac{F\pa{z_s}}^{t} - \pac{F\pa{\mu}}^{t}   }{z_{s}-  \mu } \dd \mu  }^{p_{s \, t}}
  }_{z_{\ell}= \sg_{\ell} q }
\\
=  \sul{k=0}{N-2} \paf{i g_+' }{p'_+}^k \paf{i g_-' }{p'_-}^{N-k-2}
\Dp{\ga}^n \wt{O}_{N,\, k}\pac{\sg \nu}_{\mid_{\ga=0} }  \; .
\label{sum-rule oscillante}
\end{multline}
The only difference in the proof of the oscillating sum-rule lies in the specialization of the variables $z_{\ell}$ to $\sg_{\ell} q$.
For a maximal number $N-2$ of derivatives in $I_n^{\pa{N,m}}$ there can be only one jump in the sequence $\sg_{\ell}$ \eqref{definition somme Sigma}.  Moreover,
the sequence $\eps_{s}$ has to fulfill $\Sg \eps_s p_{\sg_{s}}=m\pa{p_+-p_-}$. It thus follows that
$\Sg \eps_s p_{\sg_{s}}=m\pa{g_+-g_-}$ and the resulting prefactor cancels with the one appearing
\eqref{expression term oscillant ordre n}.

\section{The action on $x$-dependent functions}

We will now apply the sum-rules derived in the previous section so as to obtain the $\e{O}\pa{1}$ part of the action of the
functionals $I_n^{\pa{N;m}}$, $m\in \paa{0,\pm 1}$, on  a certain class of $x$-dependent functions. We however first  define
the class of functions that we consider in the following.
\subsection{The type of $x$-dependent function}
In the rest of this article, we assume that $\mc{F}_n$ has the typical structure of functions involved
in the asymptotic analysis of correlation functions of integrable models \cite{KozKitMailSlaTerXXZsgZsgZAsymptotics}.
Such functions $\mc{F}_n$ are written as
\beq
\mc{F}_n\pab{ \paa{\la} }{ \paa{z} } =
\exp\paa{ x \mc{G}_n\pab{ \paa{\la} }{ \paa{z} } + \ln x \, \mc{H}_n\pab{ \paa{\la} }{ \paa{z} }   }
\mc{W}_n\pab{ \paa{\la} }{ \paa{z} }  \pl{i=1}{n} \mc{V}_n\pamb{\la_i} {\paa{\la} }{ \paa{z} } \; .
\label{definition x dependent functions}
\enq
All of the function in \eqref{definition x dependent functions} are symmetric in the $\paa{\la}$ and $\paa{z}$ variables. Moreover, they satisfy reduction properties
in respect to variables appearing in hypergeometric notation, \textit{eg}:
\beq
\mc{W}_n\pab{ \paa{\la}_1^n }{ \paa{z}_1^n } _{\mid_{z_s=\la_k }} =
\mc{W}_n\pab{ \paa{\la}_{1, \, \not= k} ^n }{ \paa{z}_{1 ,\,  \not =s}^n }  \; .
\label{definition reduction W}
\enq
In the above equation, we have indicated the label dependence of the variables
so as to make clear what we mean by reduction property. Yet, as long as it does not lead to confusion, we will omit to write this dependence explicitly later on.
\subsection{The action of $I_n^{\pa{N,0}}$}

\begin{prop}
Let $\mc{F}_n$ be as in \eqref{definition x dependent functions}, then
\beq
x^{-N}I_n^{ \pa{N;0} }\pac{\mc{F}_n} =  \ex{x \mc{G} + \ln x \mc{H} } W \paa{\f{\pa{-1}^{n-1}}{ \pa{n-1}! } \Dp{\ga}^n \pa{\nu_0}_{\mid_{\ga=0}}  } \sul{\sg=\pm }{} \f{1}{N}\paf{i \mc{G}'\pa{\sg q}}{p'\pa{\sg q}}^N
\mc{V}^N\pa{\sg q} \; +\e{o}\pa{1} \;.
\enq
We have introduced shorthand notations for the fully reduced functions
\beqa
\mc{G}=\mc{G}_0\pab{ \cdot}{ \cdot  } = \mc{G}_1\pab{z}{z} \; , \quad \mc{H}=\mc{H}_0\pab{ \cdot}{ \cdot  }   \; ,
\quad \mc{W}=\mc{W}_0\pab{ \cdot}{ \cdot  } \; ,  \\
\quad \mc{V}\pa{\la}=\mc{V}\pamb{\la}{ \cdot}{ \cdot  }  \; \;\; , \qquad
\quad \mc{G}'\pa{\la}= \Dp{\eps}\mc{G}_1\pab{ \la }{\la+\eps} _{\mid_{\eps=0}} \; .
\eeqa
\end{prop}

\Proof

The key observation is that for functions satisfying reduction properties, may terms in the action of $I_{n}^{\pa{N,m}}$
decouple and one is almost reduced to the pure-product case.
Due to the reduction properties of $\mc{F}_n$, the action of the finite-difference operators simplifies:
\bem
\eth^{\pa{t}}_{z}\pa{\mu_{\ell, \, t, \, j}}  \cdot \mc{F}_n\pab{ \paa{z_s}_{}^{\ov{p}_s} }{ \paa{z_s}^{\ov{p}_s+\eps_s}  }=
\mc{F}_{n-t}\pab{ \paa{z_s}_{}^{\ov{p}_s-t\de_{s, \, \ell}} }{ \paa{z_s}^{\ov{p}_s+\eps_s-t\de_{s , \, \ell} }  }
\left\{ \pac{ \mc{V}_{n-t}} ^t \pamb{ z_{\ell} }{  \paa{z_s}_{}^{\ov{p}_s-t\de_{s ,\, \ell}}   } {   \paa{z_s}^{\ov{p}_s+\eps_s-t\de_{s, \, \ell} }   } \; \right. \\
 \left. - \pac{\mc{V}_{n-t} } ^t \pamb{ \mu_{\ell, \, m, \, j}  }{  \paa{z_s}_{}^{\ov{p}_s-t\de_{s, \, \ell}}   } {   \paa{z_s}^{\ov{p}_s+\eps_s-t\de_{s, \, \ell} }   }
\right\} \; .
\end{multline}
 There, $\de_{s,\ell}$ is the Kronecker $\de$ symbol. Thus,
\bem
\pl{\ell=1}{r} \pl{t=1}{n} \pl{j=1}{p_{\ell\, , t}}
\paa{   \Int{-q}{q}  \f{\dd \mu_{\ell, \, t, \, j} }{z_{\ell}-  \mu_{\ell, \, t, \, j} } \eth^{\pa{t}}_{z}\pa{\mu_{\ell, \, t, \, j}} }
\cdot \mc{F}_n\pab{ \paa{z_s}_{}^{\ov{p}_s} }{ \paa{z_s}^{\ov{p}_s+\eps_s}  } \\
=\mc{F}_{\abs{p} } \pab{ \paa{z_s}^{p_{s\, 0}} }{ \paa{z_s}^{p_{s\, 0}+\eps_s} } \pl{\ell=1}{r} \pl{t=1}{n}
\paa{ \Int{-q}{q} \dd \mu   \f{ \pac{ \mc{V}_{\abs{p}} }^t \pamb{z_{\ell}}{  \paa{z_s}^{p_{s\, 0}}  }{ \paa{z_s}^{p_{s\, 0}+\eps_s}  } -
\pac{\mc{V}_{\abs{p}} }^t\pamb{\mu}{  \paa{z_s}^{p_{s\, 0}}  }{ \paa{z_s}^{p_{s\, 0}+\eps_s}  }  }
{z_{\ell} - \mu}  } ^{p_{\ell \, t}} \; .
\label{resultat action op Diff Finie}
\end{multline}
$\abs{p}=\Sg_{\ell} p_{\ell \, 0}$ is the number of variables remaining after computing the action
the finite-difference operators. Indeed, all the dependence on the variables $\paa{z_s}$ that are repeated $p_{s, \, \ell }$ times, $\ell=1, \dots,n$, disappears from the function.
To evaluate the action of $I_n^{\pa{N,0}}$, one should hit with the product of $z_s$ derivatives on the RHS of \eqref{resultat action op Diff Finie}
just as it was done in \eqref{action derivée multiple fonction exposant}. Clearly, such an operation generates positive powers of $x$ (by differentiating the exponent $\ex{x \mc{G}_n}$) or of $\ln x$ (by differentiating  $\ex{\ln x \, \mc{H}_n}$). In order to obtain the highest possible power of $x$, the derivative should
only hit on the ${x \mc{G}_n}$ part of the exponent. In this way $\Dp{z_1}^{k_1} \dots \Dp{z_r}^{k_r}$ produces a contribution of order
$x^{k_1+\dots + k_r}$. Hence, the largest possible power of $x$ comes from derivatives such that $\Sg k_{\ell}=N$, and then
\beq
\Dp{z_1}^{k_1} \dots \Dp{z_r}^{k_r}
%
\mc{F}_{\abs{p}} \pab{ \paa{z_s}^{p_{s\, 0}} }{ \paa{z_s}^{p_{s\, 0}+\eps_s} }   =
x^N \pl{s=1}{r}  \pa{\Dp{z_k} \mc{G}_{\abs{p}} \pab{  \paa{z_s}^{p_{s\, 0}}  }{ \paa{z_s}^{p_{s\, 0}+\eps_s}  } }^{k_s}
 \mc{F}_{\abs{p}} \pab{ \paa{z_s}^{p_{s\, 0}} }{ \paa{z_s}^{p_{s\, 0}+\eps_s} }
+ \e{O}\paf{ x^N\ln x}{x} \; .
\label{Action derivée fonction x-dependent}
\enq
Note that the positive power $x^N$ simplifies with the prefactor in front of $I_n^{\pa{N,m}}$. Once the derivatives are computed, one should send all of the $z$'s to $\sg q$, $\sg=\pm$.
Hence, due to the conditions $\Sg \eps_{\ell}=0$ the upper and the lower variables in hypergeometric notation will be equal.
 This allows one to apply the reduction properties to $\mc{F}_{\abs{p}}$, as well as to most of the variables appearing in
 $\Dp{z_k} \mc{G}_{\abs{p}}$. In particular,  the reduction properties imply that
\beq
\Dp{z_k} \mc{G}_{\abs{p}} \pab{  \paa{z_s}^{p_{s\, 0}}  }{ \paa{z_s}^{p_{s\, 0}+\eps_s}  }= \eps_k
\Dp{\eps} \mc{G}_{\abs{p}} \pab{  \paa{z_s}_{s\not=k}^{p_{s\, 0}} \; ; \; z_k  }{ \paa{z_s}_{s\not = k}^{p_{s\, 0}+\eps_s} \; ; \; z_k +\eps  } _{\mid_{\eps=0}} \; .
\enq
Hence, after implementing most of the operations, it follows that, up to $\e{o}\pa{1}$ corrections,  $\mc{W}\, \ex{ x \mc{G} + \ln x\,  \mc{H}}$ can be pulled out from
the sum as it is a constant, and
\bem
x^{-N}I_n^{ \pa{N,0} }\pac{\mc{F}_n}= \mc{W}\ex{x \mc{G} + \ln x \mc{H} } \mf{S}^{\pa{m}}_N * \sul{ \Sg k_{\ell} = N }{}
G^{\mf{S}^{\pa{m}}_N}_{\paa{k_{\ell}}}  \\
\times \paa{  \pl{s=1}{r} \pac{ \eps_s \mc{G}'\pa{z_s} }^{k_s}   \pac{\mc{V}\pa{z_s} }^{p_{s\,0}}
\pl{t=1}{n}  \paa{ \Int{-q}{q}  \dd \mu  \f{ \pac{  \mc{V}\pa{z_s} }^t-\pac{ \mc{V}\pa{\mu} }^t }{ z_s-\mu }  }^{p_{\ell \, t}}  } + \e{o}\pa{1}
%
%
%
%
\end{multline}
Thus by applying the sum-rule \eqref{sum-rule non-oscillating} we get the claim. \qed
%
%
%
%
%
%
%
%
%
%

\subsection{The action of $I_n^{\pa{N,\pm 1}}$}

We now focus on the action of the functional $x^{-N}I^{\pa{N,m}}_n$, $m=\pm$ and $N\geq 2$, on the class of $x$-dependent functions as in \eqref{definition x dependent functions}.

\begin{prop}
Let $\mc{F}_n$ be as in \eqref{definition x dependent functions}, then
\bem
\f{I_n^{ \pa{N;m} }\pac{\mc{F}_n}}{x^{N}}= x^{-2}\mc{W}^{\pa{m}}\ex{x \mc{G}^{\pa{m}} + \ln x \mc{H}^{\pa{m}} }  \sul{k=0}{N} \paf{i \pa{\mc{G}^{\pa{m}}}'_+ }{p'_+}^k \paf{i \pa{\mc{G}^{\pa{m}}}'_- }{p'_-}^{N-k}
\Dp{\ga}^n \wt{O}_{N,\, k}\pac{\sg \nu^{\pa{m}}}_{\mid_{\ga=0} }  \; +  \; \e{o}\pa{x^{-2}} \\
\; \e{with} \quad \nu^{\pa{m}}\pa{z}=\f{i}{2\pi} \ln \pa{ 1+ \ga \mc{V}^{\pa{m}}\pa{z} } \; .
\end{multline}
There we have introduced a notation for the almost reduced functions
\bem
\mc{W}^{\pa{m}}=\mc{W}_1\pab{-mq}{mq} \; ,\quad \mc{H}^{\pa{m}}\pa{z}=\mc{H}_1\pab{-mq}{mq} \; ,
\quad \mc{V}^{\pa{m}}\pa{z}=\mc{V}_1\pamb{z}{-mq}{mq} \; , \\
 \e{as} \,  \e{well} \, \e{as}  \qquad \pa{\mc{G}^{\pa{m}}}'\pa{z}=\Dp{\eps} \mc{G}_2\pab{ -mq, z }{mq , z+\eps} \; . \hspace{5cm}
\end{multline}
\end{prop}

\Proof
If one restricts oneself to the leading $\e{O}\pa{1}$ part of this action, then, as before,
one should only consider the part of the sum corresponding to integers $k_{\ell}$ such that $\sul{}{}k_{\ell}=N-2$. All the others will produce subdominant contributions.
Next, one should once again act with the derivatives on the exponent $\exp\paa{x \mc{G}_n}$ exactly as it was done in \eqref{Action derivée fonction x-dependent}.

Note that, for the oscillating functional we consider, there are at least $2$ variables $z_{\ell}$. Moreover, after the computation of derivatives one should set $z_1=\dots=z_p=\sg q$ and $z_{p+1}=\dots=z_r=-\sg q$, $p \in \intff{1}{ r-1}$. Indeed, for the slowest oscillating terms when a maximal number
of derivatives $\Sg k_{\ell}=N-2$ is considered, then there is only  one jump in the sequence $\sg_{\ell}$.

The first step of the analysis, that is to say
the determination of the action  of the $\eth^{\pa{t}}_{z_{\ell}}\pa{\mu_{\ell, \, t, \, j}} $ operators is the same as in the non-oscillating case.
Then, in order to recover an $\e{O}\pa{x^{-2}}$ contribution, one should act with all of the derivatives  only on the exponent $x \mc{G}_n$. The only difference in the
proof consists in  to the specifications of the $z$'s. It is not difficult to see that, for any function satisfying the reduction properties
\eqref{definition reduction W}:
\beq
\mc{W}_r \pab{ \paa{\sg q}^p \cup \paa{-\sg q}^{r-p} }{ \paa{\sg q}^{p+\sg m} \cup \paa{-\sg q}^{r-p-\sg m}  }=
\mc{W}_1 \pab{ -m q }{ m q} \; .
\enq
Thence,
\bem
x^{-N} I_n^{ \pa{N;m} }\pac{\mc{F}_n}= x^{-2}\mc{W}^{\pa{m}}\ex{x \mc{G}^{\pa{m}} + \ln x \mc{H}^{\pa{m}} } \;\;
 \mf{S}^{\pa{m}}_N \;* \!\! \sul{ \Sg k_{\ell} = N-2 }{} \\
\hspace{2cm} G^{ \mf{S}^{\pa{m}}_N  }_{ \paa{k_{\ell}} }
\paa{  \pl{s=1}{r} \pac{  \eps_s \mc{G}'\pa{z_s} }^{k_s}  \pac{\mc{V}^{ \pa{m} }\pa{z_s}}^{p_{s\,0}}
\pl{t=1}{n}  \paa{ \Int{-q}{q}  \dd \mu  \f{ \pac{\mc{V}^{\pa{m}}\pa{z_s} }^t- \pac{\mc{V}^{\pa{m}}\pa{\mu}}^t }{ z_s-\mu }  }^{p_{\ell \, t}}  } + \e{o}\pa{1}\;.
\end{multline}
 It is now clear that the above series is exactly of the type considered in the sum-rule \eqref{sum-rule oscillante}.  \qed

\section{Conclusion}
In this article, we have given the explicit proof of the action of the functionals involved in the
AE of cyclic integrals. The form of this action was used in \cite{KozKitMailSlaTerXXZsgZsgZAsymptotics}
for the computation of the asymptotic behaviour of the generating  function of the spin-spin correlation functions in integrable models.
Our result was proven thanks to a precise determination of the structure of a part of the AE of $\ln\ddet{}{I+V}$
to any order in $x^{-N}$. These were obtained thanks to several equivalent ways of representing the kernel
of the generalized sine kernel. We have only derived this action up to the first non-vanishing terms. However, the method that we have developed
can also be applied, with the price of increasing complexity, to compute the sub-leading corrections to these terms. These, in turn
would allow one to access to corrections in the long-distance asymptotic behavior of $\moy{\ex{\be \mc{Q}_x}}$
that deviate from the conformal field  theory  predictions.

\section*{Acknowledgements}
Karol K. Kozlowski is supported by the ANR program GIMP
ANR-05-BLAN-0029-01 and by the French ministry of research. The author  would like to thank the Centre de
recherche mathématiques of the Université de Montréal for its hospitality during the conference "Integrable quantum systems and solvable
statistical models".
The author is grateful to N.-Kitanine, J.-M. Maillet, N. A. Slavnov and V. Terras for
many stimulating discussions.

\end{document}